\documentclass[twocolumn,aps,prb,showpacs,groupedaddress]{revtex4-1}
\usepackage[latin9]{inputenc}
\usepackage{float}
\usepackage{amsmath}
\usepackage{graphicx}

\makeatletter
\usepackage{pifont}
\usepackage{amsfonts}
\usepackage{subfigure}
\usepackage{booktabs}
\usepackage{setspace}
\usepackage{threeparttable}
\usepackage{enumerate}
\usepackage{wasysym}
\usepackage{xspace}
\usepackage[T1]{fontenc}
\usepackage{textcomp}
\usepackage{txfonts}
\usepackage{bm}
\usepackage{graphicx,color,epsfig}
\usepackage{soul}

\makeatother

\begin{document}

\title{Spin-orbit interaction driven dimerization in one dimensional frustrated magnets}

\author{Shang-Shun Zhang$^{1}$, Nitin Kaushal$^{1}$, Elbio Dagotto$^{1,2}$, and Cristian D. Batista$^{1,3}$}

\address{$^1$Department of Physics and Astronomy, University of Tennessee, Knoxville,
Tennessee 37996-1200, USA}
\address{$^2$Materials Science and Technology Division, Oak Ridge National Laboratory, Oak Ridge, Tennessee 37831, USA}
\address{$^3$Quantum Condensed Matter Division and Shull-Wollan Center, Oak Ridge National Laboratory, Oak Ridge, Tennessee 37831, USA}

\begin{abstract}
We study the  effect of spin-orbit interaction on one-dimensional U(1)-invariant frustrated magnets with dominant 
critical nematic fluctuations. The spin-orbit coupling explicitly breaks the U(1) symmetry of arbitrary global spin rotations about 
the high-symmetry axis down to $Z_2$  (invariance under a $\pi$-rotation). Given that the nematic order parameter
is invariant under a $\pi$-rotation, it is relevant to ask if other discrete symmetries can be spontaneously broken.
Here we demonstrate that the spin-orbit coupling  induces a bond density wave
that spontaneously breaks the translational symmetry and opens a gap in the excitation spectrum. 
\end{abstract}
\maketitle

\section{Introduction}
Frustrated magnetism is a continuous source of exotic states of matter
that challenge the existing characterization probes~\cite{ref:review1,ref:review2}. Once quantum fluctuations melt the traditional magnetic long-range
order, it often happens that the remaining liquid or multipolar orderings do not couple directly to the usual experimental
probes.   
A simple example is  the  spin nematic phase proposed to be the ground 
of  the one-dimensional (1D) $J_1-J_2$ Heisenberg model near its saturation field~ \cite{ref:nematic1,ref:nematic2,ref:nematic3,ref:nematic4,ref:nematic5,ref:nematic5b,ref:nematic5c,ref:nematic5d,ref:nematic6,
ref:nematic7,ref:nematic8,ref:nematic9,ref:nematic10,ref:nematic11}. This phase arises from a Bose-Einstein condensation 
of magnon pairs  right below the saturation field $h_{\rm sat}$~\cite{ref:nematic1,ref:nematic2,ref:nematic3,ref:nematic4,ref:nematic5,ref:nematic5b,ref:nematic5c,ref:nematic5d}. The attractive magnon-magnon interaction is generated by a ferromagnetic (FM) nearest neighbor (NN) exchange ($J_1<0$),
which competes against  an antiferromagnetic (AFM) next-nearest-neighbor (NNN) exchange $J_2>0$. 
$J_2$ must be bigger than $|J_1|/4$ for the zero field ground state not to be ferromagnetic.

Several quasi-1D materials are approximately described by the  $J_1-J_2$ 
model with  FM and AFM exchange interactions  $J_1$ and $J_2$, respectively.  
Known examples include Rb$_2$Cu$_2$Mo$_3$O$_{12}$,~\cite{ref:1DmaterialA1,ref:1DmaterialA2}, LiCuVO$_4$,~\cite{ref:1Dmaterial2,ref:1Dmaterial3,ref:1Dmaterial4,ref:1Dmaterial6,LiCuVO4,LiCuVO4b,ref:1Dmaterial7,ref:1Dmaterial8,ref:1Dmaterial9,ref:1Dmaterial10,ref:1Dmaterial11} LiCuSbO$_4$,~ \cite{LiCusbO4,grafe2016} PbCuSO$_4$(OH)$_2$,~\cite{PbCuSO4OH2}  which span a wide range of  $J_2/\rvert J_1\rvert$ values.
However, a direct experimental observation of the predicted nematic ordering is rahter challenging.~\cite{sato09,sato11} Given the symmetry of 
the order parameter, the nematic spin ordering is expected to induce a local quadrupolar electric moment 
via the always present spin-orbit coupling combined with the lattice anisotropy. However, \emph{ the spin anisotropy induced by  this combination has the additional effect 
of breaking the global  U(1) symmetry of spin rotations along the magnetic field axis down to a finite group}.
For most of the known compounds, this group is not bigger than $Z_2$ for any direction of the applied magnetic field
(only C$_2$ rotation axes). This fact raises another concern because the nematic order parameter $\langle S_r^{+}S_{r+1}^+\rangle$ does not break the remaining $Z_2$ group. In other words, the nematic order parameter is invariant under $\pi$-rotations. 
This simple observation implies that if some form of ordering still exists right below   the high field paramagnetic phase of these compounds, it should not be called ``nematic ordering''.
Nevertheless, the dominant nematic susceptibility of the U(1) invariant model may still induce additional symmetry breaking in the presence of spin anisotropy. If this is the case, it is important to identify those discrete symmetries.

\begin{figure}[!t]
\includegraphics[scale=0.35]{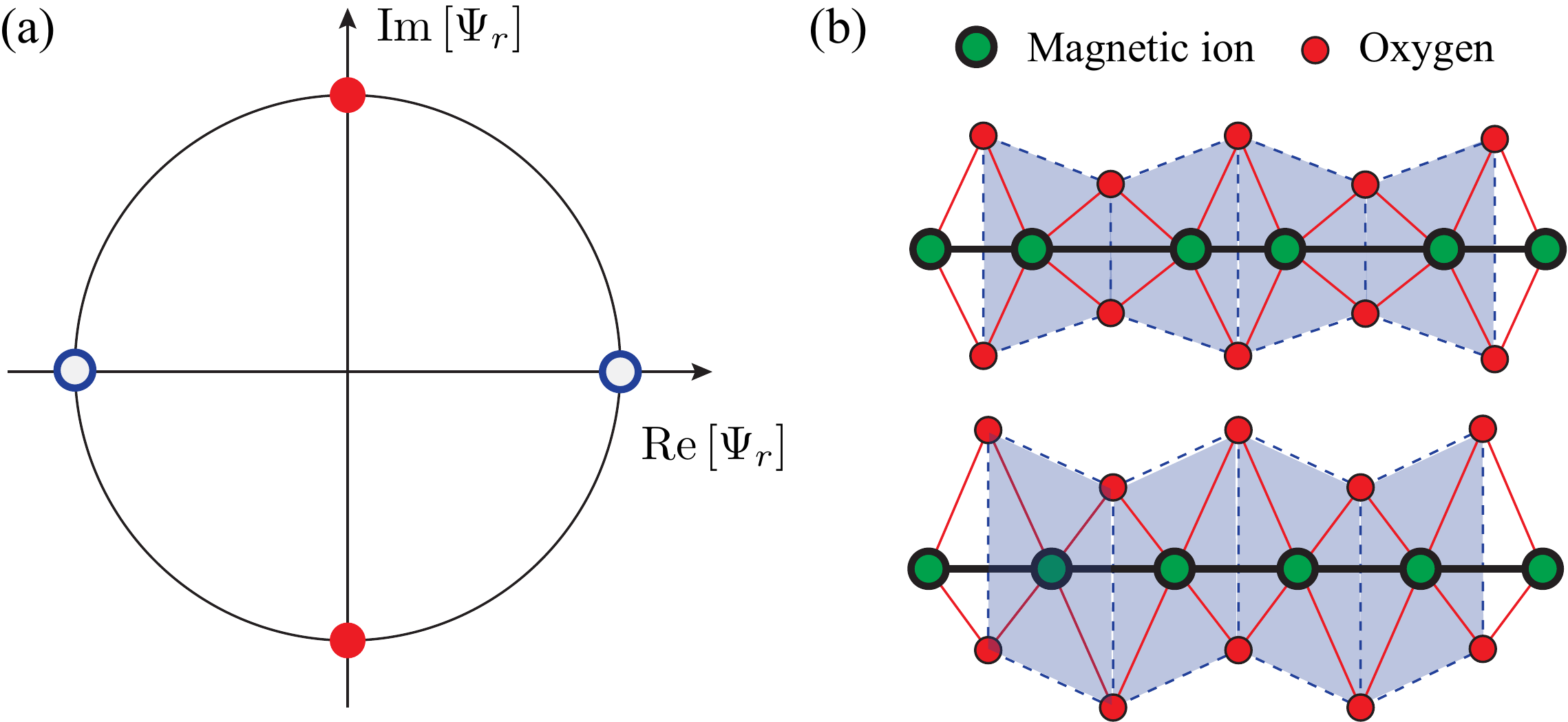}
\caption{(Color online) (a) The two scenarios of nematic bond order parameter 
$\Psi_{r}=\langle S_{r}^{+}S_{r+1}^{+}\rangle = \langle {\cal O}^{a} \rangle + i \langle  {\cal O}^{b} \rangle$:
open and full circles represent the real, $\langle {\cal O}^{a} \rangle$, and imaginary, $\langle {\cal O}^{b} \rangle$ parts of the nematic order parameter, respectively. (b) Lattice distortions
induced via spin-orbit coupling by the real (upper panel) and the imaginary (lower panel) parts of the 
bond nematic order parameter. Translational symmetry is broken in both cases, but the lattice distortion
takes place along different directions. The bigger  circles represent the
magnetic transition-metal ions. The smaller circles represent the oxygen atoms that mediate the super-exchange interaction.}
\label{fig:dimerization}
\end{figure}

In this work we investigate the relevant effect of spin-orbit interaction on the 1D frustrated $J_1-J_2$ model.  
Based on our previous considerations, there are two possible scenarios:
i) The field-induced transition from a quantum paramagnet to the nematic phase is replaced by  a crossover (no
discrete symmetry breaking);
ii) The field-induced transition from a quantum paramagnet to the nematic phase is replaced by a discrete symmetry  breaking.
We can anticipate that the problem under consideration belongs to the second case because 
it is known that the magnon-pairs condense at a finite wave vector $\pm Q$. In other words, the nematic 
ordering breaks the translational symmetry, which is not affected by the inclusion of the spin-orbit interaction. 
In addition, the system has an additional $Z_2$ symmetry besides the $\pi$-rotation about the $z$-axis. This  symmetry is the product of two operations, ${\cal T} {\cal R}(\pi)$, where
${\cal T}$ is the time reversal operator and ${\cal R}(\pi)$ is a $\pi$-rotation operator about an axis perpendicular to
the  field  direction. As we explain below, the real part of the nematic order parameter, ${\cal O}^a$, 
preserves this $Z_2$ symmetry, while the imaginary part, ${\cal O}^b$, does not. Fig.~\ref{fig:dimerization}~(a) shows the 
the different broken symmetry states associate with the stabilization of the real or imaginary part of the nematic order parameter.

In the following sections, we derive a phenomenological Ginzburg-Landau theory that is complemented by
microscopic analytical and numerical calculations. Moreover, we demonstrate that the combination of
a divergent nematic susceptibility and  spin anisotropy stabilizes the real component of the nematic order parameter, ${\cal O}^a$, which in turn produces  bond dimerization in most of the phase diagram. This bond ordering is accompanied by an
orthorhombic distortion of the surrounding oxygen octahedron, as it is schematically shown 
in the upper panel of  Fig.~\ref{fig:dimerization}~(b). In contrast, the  imaginary component of the  ``nematic'' order parameter, ${\cal O}^b$, does not produce bond dimerization.  If stabilized by other mechanisms, this phase  would produce a local orthorhombic distortion of the surrounding oxygen octahedron along diagonal directions, as illustrated in the lower panel of Fig.~\ref{fig:dimerization}~(b). 

Given that bond dimerization couples to the lattice via magneto-elastic interaction and lowers the space group of the material under consideration, the combined effect of high spin nematic susceptibility and spin anisotropy can in principle be detected with X-rays. In addition, the incommensurate bond-density wave expected for smaller values of $J_2/|J_1|$ should lead to a double-horn shape of the nuclear magnetic resonance (NMR) line. These conclusions can shed light on
the search for the spin ``nematic ordering'' predicted on the basis of a U(1) invariant $J_1-J_2$ Heisenberg model. 

The structure of this paper is as follows. In Sec. II, we introduce a simple model Hamiltonian in which the U(1) symmetry is reduced to Z$_2$ via the inclusion of an Ising term (symmetric anisotropy). In Sec. III we consider a simple Ginzburg-Landau (GL) theory,
which describes the possible scenarios.
The phenomenological input parameters of the GL theory are calculated in Sec. IV by means of an analytical approach to the  microscopic Hamiltonian. The results of this analytical approach are confirmed by numerical Density Matrix Renormalization Group (DMRG) calculations presented in Sec. V. The general implications of our results 
for experimental studies of unconventional magnetic ordering in quasi-1D frustrated compounds are discussed in Sec. VI.

\section{Model Hamiltonian}

We consider a spin-$1/2$ anisotropic Heisenberg model on a 1D chain
with ferromagnetic nearest-neighbor exchange, $J_{1}<0$, and antiferromagnetic
next nearest neighbor exchange $J_{2}>0$:
\begin{align}
\mathcal{H} & =J_{1}\sum_{j}{\bf S}_{j}\cdot{\bf S}_{j+1}+J_{2}\sum_{j}{\bf S}_{j}\cdot{\bf S}_{j+2}-h\sum_{j}S_{j}^{z} \nonumber \\
 & \;\;+\alpha\sum_{j}S_{j}^{x}S_{j+1}^{x},\label{eq:model}
\end{align}
The last term is an Ising interaction that in real materials arises from  the combined effect of spin-orbit coupling and lattice anisotropy. This term reduces the  U(1) symmetry of continuous spin rotations about the field axis to $Z_2$.
We note that the U(1) symmetry is restored if the magnetic field is applied along the $x$-direction. However, in  real quasi-1D materials, the U(1) symmetry is not present for any direction of the magnetic field because of the interaction with other chains \cite{ref:nematic1}. In spite of these considerations, the U(1) invariant model has been invoked  to describe various quasi-1D transition metal compounds.~\cite{sato13} 

Based on the U(1) invariant model ($\alpha=0$), several authors  proposed
that nematic quasi-long range ordering should be observed right below the saturation field $h_{\rm sat}$~\cite{ref:nematic1,ref:nematic2,ref:nematic3,ref:nematic4,ref:nematic5,ref:nematic5b,ref:nematic5c,ref:nematic5d}.  
$h_{\rm sat}$ is finite only for $J_2 > |J_1|/4$ because the zero field ground state is ferromagnetic for
$J_2 \leq |J_1|/4$. The nematic ordering corresponds 
to a Bose-Einstein condensation of two-magnon bound states with a local order parameter
$\langle S_{j}^{+}S_{j+1}^{+}\rangle$.
The attractive magnon-magnon interaction is provided by the ferromagnetic nearest-neighbor exchange $J_1$.
The  ratio $ J_{2}/\lvert J_{1} \rvert$  controls the total momentum $\pm Q$ of the two-magnon bound state. $Q$ is incommensurate for small values of $ J_{2}/\lvert J_{1} \rvert$ and it is equal to  $\pi$ for $ J_{2}/\lvert J_{1} \rvert \geq 0.375$. 

In general,  the continuous SU(2) symmetry of the Heisenberg interaction is broken down to a discrete symmetry group in real materials~\cite{ref:1DmaterialA2,ref:1Dmaterial2}. Even for an idealized single-chain system, the exchange interaction turns out to be anisotropic, instead of SU(2) invariant, once the spin-orbit interaction is included. This is so because an isolated chain has only one symmetry axis  parallel to the chain direction ($x$-direction in our notation). In other words, the exchange interaction between spin components parallel to the chain direction is different from the exchange interaction between the spin components perpendicular to the chain direction, as it is clear from the $\alpha$-term in Eq.~\eqref{eq:model}. Consequently, the pure 1D Hamiltonian has only discrete point group symmetries if the external magnetic field is not parallel to the chain direction. 
For the case under consideration (magnetic field perpendicular to the chain direction),  the U(1) symmetry of 
the Heisenberg model is reduced to a discrete  $Z_{2}$
symmetry corresponding to a $\pi$-rotation about the $z$-axis ${\cal R}_{z}(\pi):\;S_{r}^{z}\rightarrow S_{r}^{z},S_{r}^{x,y}\rightarrow-S_{r}^{x,y}$. Correspondingly, the spin components
$S_{r}^{\pm}$ transform into $e^{\pm i\phi} S_{r}^{\pm}$ under a rotation by $\phi$ about the 
$z$-axis. This means that the nematic order parameter, $\langle S_{r}^{+} S_{r}^{+}  \rangle$, transforms 
into $ e^{i 2 \phi} \langle S_{r}^{+} S_{r}^{+}  \rangle$, implying that it
is invariant under $\pi$-rotations, as expected for a director. The inclusion of spin anisotropy
then forces us to reconsider the problem because other symmetries (different from rotations)  have to be invoked  to characterize the phase that replaces the nematic quasi-long range ordering.

Besides the above mentioned $\pi$ rotation about the $z$-axis, the Hamiltonian of Eq.~\eqref{eq:model}
is invariant under the product of  the time reversal operation and a $\pi$-rotation about the $y$-axis,
${\cal T}{\cal R}_{y}(\pi)$, which changes the sign of the
$y$ spin component: $S_{r}^{x,z}\rightarrow S_{r}^{x,z},S_{r}^{y}\rightarrow-S_{r}^{y}$. 
The real-part of the nematic order parameter,
\begin{equation}
\Re\langle S_{r}^{+}S_{r+1}^{+}\rangle=\langle S_{r}^{x}S_{r+1}^{x}-S_{r}^{y}S_{r+1}^{y}\rangle \equiv \langle {\cal O}^a (r) \rangle,
\end{equation}
remains invariant under this transformation. In contrast, the imaginary part, 
\begin{equation}
\Im\langle S_{r}^{+}S_{r+1}^{+}\rangle=\langle S_{r}^{x}S_{r+1}^{y}+S_{r}^{y}S_{r+1}^{x}\rangle,
\end{equation}
changes sign. Finally, the nematic order parameter breaks the translational symmetry because the magnon-pairs condense at a finite momentum $\pm Q$. This symmetry is then expected to break spontaneously  for $\alpha \neq 0$, as long as $Q$ is commensurate. 

Based on this simple symmetry analysis, the spin anisotropy should stabilize a state that breaks the
translational symmetry (in a strong or a weak sense) and select either the real or the imaginary part of the original nematic order parameter. Only one of these two components should be selected, as supposed to some linear combination, because they belong to different irreducible representations of the point group of $\mathcal{H}$.

\section{Ginzburg-Landau theory}
\label{GG}

The attractive interaction between magnons arising from the ferromagnetic nearest-neighbor interaction, leads to  two-magnon bound states for $J_{2}/|J_{1}|>1/4$. The minimum energy of the two-magnon bound state is achieved for a finite value,  $\pm Q$, of the center of mass momentum. The two-magnon bound states condense for  $ h < h_{c}$ (note that $h_c$ is higher than the  field required to close the single-magnon gap). The two-magnon condensate is 
characterized by a two component complex order parameter $\Psi_{\pm Q}$ (macroscopic wave function
of condensate) whenever $Q \neq -Q$. The spin-orbit interaction generates an effective coupling between these two components, as it can be inferred from the lowest order expansion of the Ginzburg-Landau
free energy:
\begin{equation}
{\cal F}=r\left(|\Psi_{Q}|^{2}+|\Psi_{\bar{Q}}|^{2}\right)+u\left(\Psi_{Q}^{*}\Psi_{\bar{Q}}^{*}+\Psi_{\bar{Q}}\Psi_{Q}\right),
\end{equation}
where $\bar{Q}=-Q$. Due to the $Z_{2}$ symmetry restriction,
the complex field $\Psi_{\pm Q}$ is fixed up to a phase factor $\pm1$.
We have also assumed that $u$ is real based on the underlying microscopic
theory. We will first assume $Q=\pi$ which is the condensation wave vector for  $J_{2}>J_{2c}$
($J_{2c}\simeq 0.375\rvert J_1 \rvert$ for $\alpha=0$). Given that $Q=\pi$ is invariant 
under spatial inversion, we have $\Psi_{\pi}=\Psi_{-\pi}$. Then,
upon minimization of the free energy, we obtain a real order parameter $\Psi_{\pi}$  for
$u<0$, and a purely imaginary order parameter for $u>0$. In the original spin language, we have
$\Psi_{\pi}=\frac{1}{N}\sum_{r}e^{i\pi r}\langle S_{r}^{-}S_{r+1}^{-}\rangle$,
whose real and imaginary parts are
\begin{align}
{\cal O}^{a} & =\sum(-1)^{r}\langle S_{r}^{x}S_{r+1}^{x}-S_{r}^{y}S_{r+1}^{y}\rangle,\\
{\cal O}^{b} & =-\sum(-1)^{r}\langle S_{r}^{x}S_{r+1}^{y}+S_{r}^{y}S_{r+1}^{x}\rangle.
\end{align}
A real order parameter only breaks the translational
symmetry, while an imaginary order parameter breaks additional symmetries, such as, ${\cal T}{\cal R}_{y}(\pi):\;S_{r}^{x,z}\rightarrow S_{r}^{x,z},S_{r}^{y}\rightarrow-S_{r}^{y}$.
In both cases, the system should develop long-range ordering at $T=0$ because
only discrete symmetries are broken. We note that the
spin anisotropy corresponds to a uniform nematic field that couples linearly  to  the uniform component 
of the nematic order parameter $\Psi_{0}=\frac{1}{N}\sum_{r}\langle S_{r}^{-}S_{r+1}^{-}\rangle$,
implying that $\Psi_{0}$ becomes finite for a finite $\alpha$.
As we will see next, the interference between $\Psi_{0}$ and the $\pi$ component, $\Psi_{\pi}$,
of the {\it real} part of the order parameter leads to a real space modulation (dimerization) of the expectation value of
nearest-neighbor bond operators.

In general, the nematic order parameter is a complex number, $|\Psi_{\pi}|e^{i\theta}$,
where $\theta=0$ and $\theta=\frac{\pi}{2}$ correspond to ${\cal O}^{a}$
and ${\cal O}^{b}$, respectively. The real space version of these order parameters is obtained via a Fourier transformation,
\begin{align}
{\cal O}^{a}(r) & =\Psi_{0}+2(-1)^{r}\Psi_{\pi}\cos(\theta),\\
{\cal O}^{b}(r) & =2(-1)^{r}\Psi_{\pi}\sin(\theta),
\end{align}
which gives the real and imaginary parts of  $\Psi_{r}=\langle S_{r}^{-}S_{r+1}^{-}\rangle$. 
It is clear that the amplitude of the real component, ${\cal O}^{a}(r)$,
is modulated in real space for finite values of the spin-orbit interaction ($\Psi_{0} \neq 0$).
In contrast, only the phase is modulated for $\alpha=0$.
In other words, the spin-orbit coupling induces a {\it dimerized bond
ordering} if the real component of the original nematic order parameter is selected. This interference between  the $0$ and $\pi$ components of ${\cal O}^{a}$ also leads to a {\it magnon pair}  density wave:
\begin{align}
n_{pair}(r) & =\langle\Psi_{r}^{\dagger}\Psi_{r}\rangle\simeq\langle\Psi_{r}\rangle^{*}\langle\Psi_{r}\rangle\nonumber \\
 & =\Psi_{0}^{2}+4\Psi_{\pi}^{2}+4(-1)^{r}\Psi_{0}\Psi_{\pi}\cos(\theta).
\end{align}
It follows that the long range ${\cal O}^{a}$ ordering is accompanied by another bond ordering
associated with the longitudinal spin component
\begin{equation}
\langle S_{r}^{z}S_{r+1}^{z}\rangle\simeq\langle\Psi_{r}\rangle^{*}\langle\Psi_{r}\rangle+ const..
\label{mod}
\end{equation}
This is just the usual bond dimerization that appears in spin-Peierls systems.~\cite{Miller82}  Indeed, Eq.~\eqref{mod} implies that the usual bond order parameter,  $\langle{\bf S}_{r}\cdot{\bf S}_{r+1}\rangle$, must also exhibit dimerization. 

The condensation wave vector becomes incommensurate ($Q<\pi$) for smaller values of $J_{2}/|J_{1}|$ (about $0.375$ for $\alpha=0$). In this case we need to consider a two component  order parameter with phases  $e^{i\theta_{\pm}}$
for $\pm Q$. Minimization of the free energy leads to
$\theta_{+}+\theta_{-}=0\text{ mod }(2\pi)$ for $u<0$ and to
$\phi_{+}+\phi_{-}=\pi\text{ mod }(2\pi)$ for $u>0$.
The complex order parameter does not have a fixed phase because of
the additional $U(1)$ phase factor $e^{\pm iQr}$, arising from translational
symmetry. This $U(1)$ symmetry precludes long-range order for the single-chain problem.
The  free energy minimization also leads to the same amplitude for both components of the order parameter:
$|\Psi_{Q}|=|\Psi_{\bar{Q}}|$, implying that the ground state must exhibit
quasi-long range bond density wave order ${\cal O}^{a}(r)=\Psi_{0}+2|\Psi_{Q}|\cos(Qr)$
or ${\cal O}^{b}(r)=2|\Psi_{Q}|\sin(Qr)$.

\section{Microscopic theory}

\begin{figure}[!t]
\includegraphics[scale=0.8]{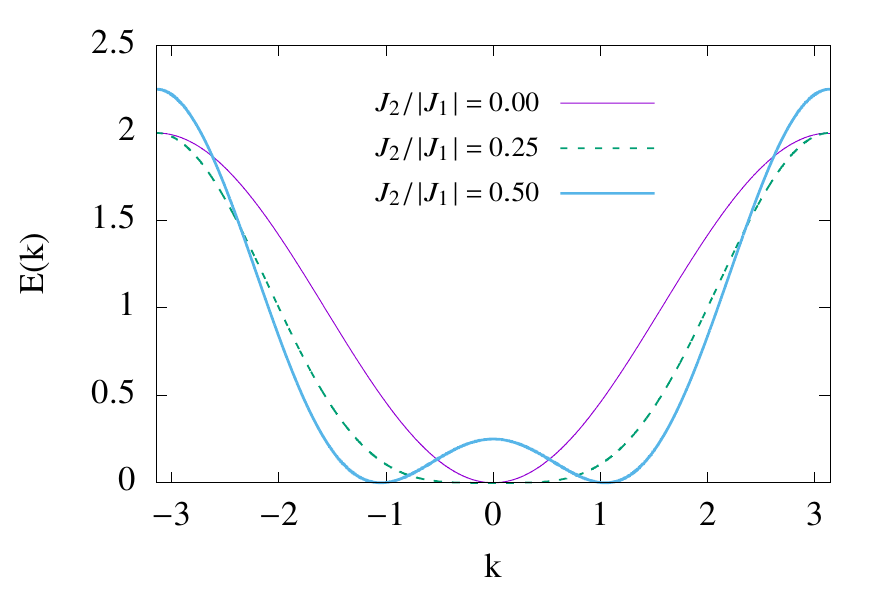}
\caption{(Color online) Single-magnon dispersion produced from the non-interacting part of the Hamiltonian [see Eq~\eqref{eq:H0}] with $\alpha=0$. The
dashed line corresponds to the Lifshitz point where the single-magnon dispersion becomes quartic at low-enrgies because  the single minimum splits into two minima.}
\label{fig:spectrum}
\end{figure}

Our discussion in  the previous section  indicates that two different kinds of bond order can be induced by spin anisotropy. To determine which order parameter is selected we need to consider the
underlying microscopic theory. 
To this end, we use the Jordan-Wigner transformation to reformulate the spin Hamiltonian (\ref{eq:model}) as a model for
 interacting spinless fermions:
\begin{align}
S_{j}^{+} & =e^{-i\pi\sum_{k=1}^{j-1}n_{k}}c_{j}^{\dagger},\\
S_{j}^{-} & =e^{i\pi\sum_{k=1}^{j-1}n_{k}}c_{j},\\
S_{j}^{z} & =c_{j}^{\dagger}c_{j}-\frac{1}{2},
\end{align}
where $c_{j}$ is the fermionic spinless operator which represents
a spin flip on site $j$. The fermionic Hamiltonian is defined as
\[
{\cal H}={\cal H}_{0}+{\cal H}_{int},
\]
where
\begin{align}\label{eq:H0}
{\cal H}_{0} & =\sum_{k}c_{k}^{\dagger}c_{k}\left(J_{1}\cos k+J_{2}\cos2k-\left(h+J_{1}+J_{2}\right)\right)\nonumber \\
 & +i\frac{\alpha}{4}\sum_{k}\sin k\left(c_{k}^{\dagger}c_{-k}^{\dagger}-c_{-k}c_{k}\right).
\end{align}
For $\alpha=0$, the single-particle spectrum, corresponding to 
single-magnon excitations, has a minimum at $k_{0}=\arccos(-\frac{J_{1}}{4J_{2}})$
if $4J_{2}>|J_{1}|$ and at $k_{0}=0$  if $4J_{2}<|J_{1}|$. Fig. \ref{fig:spectrum} shows
the evolution of the single-particle spectrum with increasing $J_2/\rvert J_1\rvert$.
The  spin-orbit interaction ($\alpha \neq 0$) breaks the $U(1)$ symmetry
and the fermion number is no longer  conserved. The interacting
part of the Hamiltonian is  
\begin{align}
{\cal H}_{int} & =\frac{1}{2N}\sum_{Q,q,p}U(Q,q,p)c_{\frac{Q}{2}+p}^{\dagger}c_{\frac{Q}{2}-p}^{\dagger}c_{\frac{Q}{2}-q}c_{\frac{Q}{2}+q},
\end{align}
where
\begin{align}
U(Q,q,p) & =2({J_{1}}+2J_{2}\cos Q)\sin(q)\sin(p)\\
 & +2{J_{2}}\sin(2q)\sin(2p).
\end{align}
The effective attractive interaction between
nearest neighbor sites, $J_{1}+2J_{2}\cos Q$, is maximized at $Q=\pi$.  Therefore, the lowest energy ``two-magnon'' bound state is
expected to have this momentum for large enough $J_2/|J_1|$.

 In the isotropic limit, the nematic phase corresponds to a magnon pair condensate.
This state can be approximated by  a coherent state built with the two-particle wave function of the bound state. In presence of magnetic anisotropy, the particle number is not conserved because particles can be created or annihilated in pairs. The Hamiltonian
eigenstates can then be grouped into two
categories based on the particle number parity. The following analysis assumes that
the magnetic anisotropy is weak ($\alpha \ll 1$), implying that the finite particle density induced by the spin anisotropy term above and at the critical field, $h_c$ can be made arbitrarily small for small enough values of $\alpha$. This condition also guarantees that
two magnon condensation is still  the dominant instability for $\alpha \neq 0$.
For this reason, we will still refer to the bound state as a ``two-magnon'' bound states (adiabatic continuation) and 
we will use the two-particle Green's function to compute the energy of the two-magnon modes.
The two-particle Green's function is obtained from the two-particle scattering amplitude by 
solving the corresponding Bethe-Salpeter equation.

\subsection{Bogoliubov representation}

The above non-interacting fermionic Hamiltonian can be diagonalized
with a  Bogoliubov transformation,
\begin{align}
c_{k} & =\left(u_{k}\alpha_{k}+v_{k}\alpha_{-k}^{\dagger}\right)e^{i\frac{\pi}{4}},\\
c_{k}^{\dagger} & =\left(u_{k}\alpha_{k}^{\dagger}+v_{k}\alpha_{-k}\right)e^{-i\frac{\pi}{4}},
\end{align}
where 
\begin{align}
u_{k} & =\sqrt{\frac{1}{2}\left(1+\frac{\epsilon_{k}}{E_{k}}\right)},\\
v_{k} & =-sign(\Delta_{k})\sqrt{\frac{1}{2}\left(1-\frac{\epsilon_{k}}{E_{k}}\right)},
\end{align}
with $E_{k}=\sqrt{\epsilon_{k}^{2}+4\Delta_{k}^{2}}$, $\Delta_{k}=\frac{\alpha}{4}\sin k$
and 
\begin{equation}
\epsilon_{k}=J_{1}\cos k+J_{2}\cos2k-\left(h+J_{1}+J_{2}\right).
\end{equation}
The diagonal Hamiltonian 
\begin{equation}
{\cal H}_{0}=\sum_{k}E_{k}\alpha_{k}^{\dagger}\alpha_{k}+E_{0}.
\end{equation}
leads to the non-interacting Green's function 
\begin{align}
iG_{0}(k,\omega) & =\langle0|{\cal T}\alpha_{k}^{\dagger}\alpha_{k}|0\rangle=\frac{i}{\omega-E_{k}+i0^{+}}.
\end{align}

The next step is to write the interaction vertex in terms of the Bogoliubov quasi-particle operators.
The normal interaction term is 
\begin{align}
{\cal H}_{int}^{N} & =\frac{1}{4N}\sum_{Q,q,p}\Gamma_{Q}^{(0)N}(q,p)\alpha_{\frac{Q}{2}+p}^{\dagger}\alpha_{\frac{Q}{2}-p}^{\dagger}\alpha_{\frac{Q}{2}-q}\alpha_{\frac{Q}{2}+q},
\end{align}
with a  normal interaction vertex \begin{widetext}
\begin{align}\label{eq:normalV}
\Gamma_{Q}^{(0)N}(q,p) & =2U(Q,q,p)\times\left(u_{\frac{Q}{2}+p}u_{\frac{Q}{2}-p}u_{\frac{Q}{2}-q}u_{\frac{Q}{2}+q}+v_{\frac{Q}{2}+p}v_{\frac{Q}{2}-p}v_{\frac{Q}{2}-q}v_{\frac{Q}{2}+q}\right)\\
 & -4U(p+q,\frac{Q-p+q}{2},\frac{Q+p-q}{2})\left(u_{\frac{Q}{2}+p}v_{\frac{Q}{2}-p}v_{\frac{Q}{2}-q}u_{\frac{Q}{2}+q}+u_{\frac{Q}{2}-p}v_{\frac{Q}{2}+p}v_{\frac{Q}{2}+q}u_{\frac{Q}{2}-q}\right).
\end{align}
We can verify that $\Gamma_{Q}^{(0)N}(q,p)=\Gamma_{Q}^{(0)N}(-q,-p)$
due to the fermionic statistics. Furthermore, $\Gamma_{Q}^{(0)N}(q,p)=\Gamma_{\bar{Q}}^{(0)N}(-q,-p)=\Gamma_{\bar{Q}}^{(0)N}(q,p)$
because of inversion symmetry. The interaction vertex has been symmetrized
with respect to the exchange of external lines.
The anomalous  interaction terms of the form $\alpha^{\dagger}\alpha^{\dagger}\alpha^{\dagger}\alpha^{\dagger}$
and $\alpha\alpha\alpha\alpha$ are
\begin{align}
{\cal H}_{int}^{A} & =\frac{1}{4!N}\sum_{Q,q,p}\Gamma_{Q}^{(0)A}(q,p)\left(\alpha_{\frac{Q}{2}+p}^{\dagger}\alpha_{\frac{Q}{2}-p}^{\dagger}\alpha_{-\frac{Q}{2}+q}^{\dagger}\alpha_{-\frac{Q}{2}-q}^{\dagger}+h.c.\right),
\end{align}
with an anomalous interaction vertex 
\begin{align}\label{eq:anomalousV}
\Gamma_{Q}^{(0)A}(q,p) & =2U(Q,q,p)\left(u_{\frac{Q}{2}+p}u_{\frac{Q}{2}-p}v_{\frac{Q}{2}-q}v_{\frac{Q}{2}+q}+v_{\frac{Q}{2}+p}v_{\frac{Q}{2}-p}u_{\frac{Q}{2}-q}u_{\frac{Q}{2}+q}\right)\\
 & -2U(p+q,\frac{Q-p+q}{2},\frac{Q+p-q}{2})\left(u_{\frac{Q}{2}+p}v_{\frac{Q}{2}-p}u_{\frac{Q}{2}-q}v_{\frac{Q}{2}+q}+v_{\frac{Q}{2}+p}u_{\frac{Q}{2}-p}v_{\frac{Q}{2}-q}u_{\frac{Q}{2}+q}\right)\\
 & +2U(p-q,\frac{Q+p+q}{2},\frac{Q-p-q}{2})\left(v_{\frac{Q}{2}+p}u_{\frac{Q}{2}-p}u_{\frac{Q}{2}-q}v_{\frac{Q}{2}+q}+u_{\frac{Q}{2}+p}v_{\frac{Q}{2}-p}v_{\frac{Q}{2}-q}u_{\frac{Q}{2}+q}\right).
\end{align}
We can verify that $\Gamma_{Q}^{(0)A}(q,p)=\Gamma_{\bar{Q}}^{(0)A}(p,q)=\Gamma_{Q}^{(0)A}(-q,-p)$
due to fermionic statistics and  $\Gamma_{Q}^{(0)A}(q,p)=\Gamma_{\bar{Q}}^{(0)A}(-q,-p)=\Gamma_{\bar{Q}}^{(0)A}(q,p)$
due to inversion symmetry. This interaction vertex has also been symmetrized
with respect to the exchange of external lines.

The remaining anomalous terms ($\alpha^{\dagger}\alpha^{\dagger}\alpha^{\dagger}\alpha$
and $\alpha\alpha\alpha\alpha^{\dagger}$) will not be considered because they give subdominant contributions (in an expansion in powers of $\alpha$) to the effective anomalous interaction vertex  shown in Fig.~\ref{fig:BSE}. We note that the nature of the order parameter is determined by the effective anomalous interaction vertex because the effective normal vertex preserves the U(1) symmetry of ${\cal H}(\alpha=0)$.

Finally, we  also note the existence of additional diagrams which are obtained by exchange of the external legs of the diagrams shown in Fig.~\ref{fig:BSE}. These diagrams are of the same order in an expansion in powers of $\alpha$. However, for the purpose of calculating the two magnon bound state, they  can be neglected because they give contributions that remain regular in the proximity
of the critical field $h=h_c$.

\begin{figure}[t!]
\centering
\includegraphics[scale=0.6]{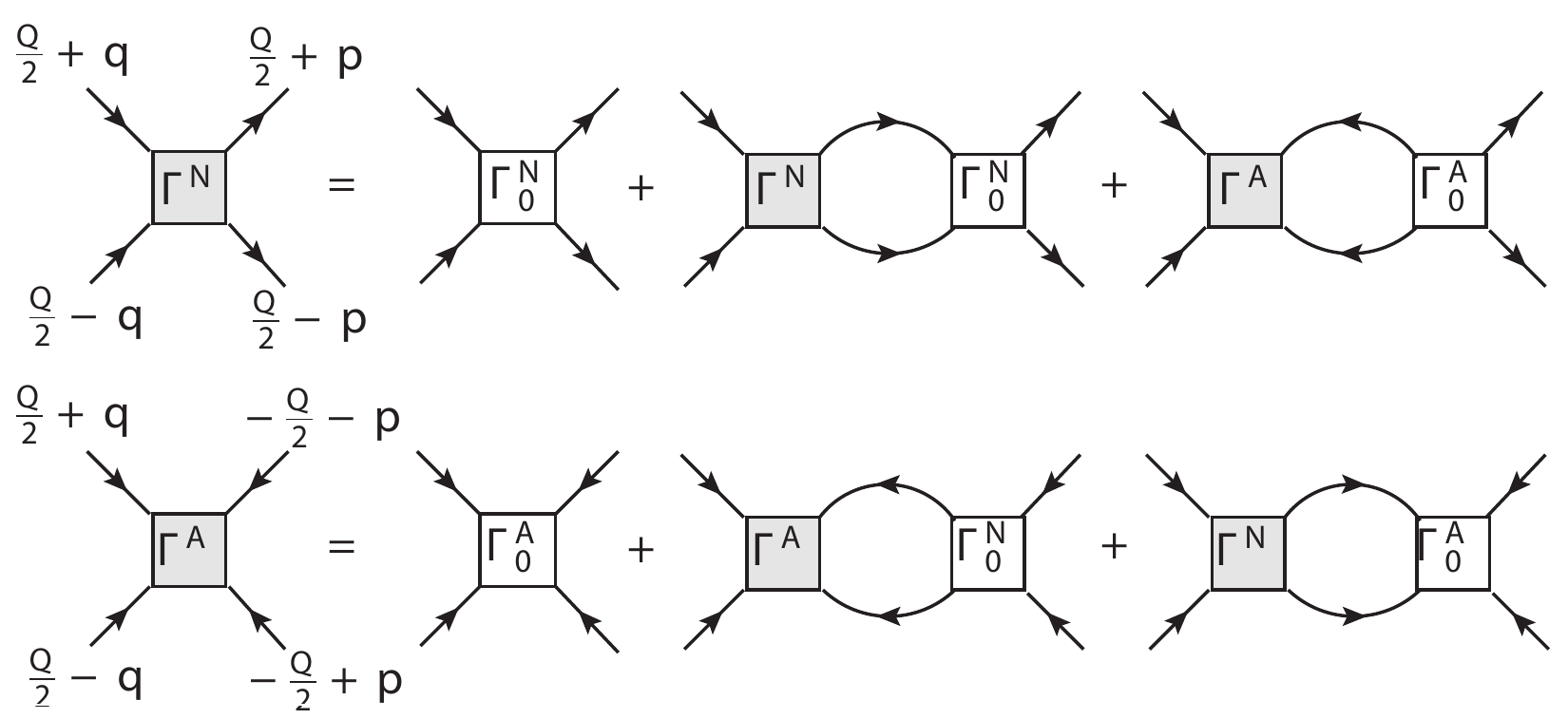}
\caption{Ladder diagrams for the normal and anomalous scattering
amplitudes $\Gamma^{N}$ and $\Gamma^{A}$.}
\label{fig:BSE}
\end{figure}

\subsection{Bethe-Salpeter equation}

In the dilute limit, the scattering amplitude can be calculated by
summing up the ladder diagrams shown in Fig. \ref{fig:BSE}. This sum leads
to the Bethe-Salpeter equation \cite{BSE}:
\begin{align}
\Gamma_{\omega Q}^{N}(q,p) & =\Gamma_{Q}^{(0)N}(q,p)-\frac{1}{2N}\sum_{k}\frac{\Gamma_{Q}^{(0)N}(Q;q,k)\Gamma_{\omega Q}^{N}(k,p)}{E_{\frac{Q}{2}+k}+E_{\frac{Q}{2}-k}-\omega-i0^{+}}-\frac{1}{2N}\sum_{k}\frac{\Gamma_{Q}^{(0)A}(q,k)\Gamma_{\omega Q}^{A}(k,p)}{E_{\frac{Q}{2}+k}+E_{\frac{Q}{2}-k}+\omega-i0^{+}},\label{eq:BSeq1}\\
\Gamma_{\omega Q}^{A}(q,p) & =\Gamma_{Q}^{(0)A}(q,p)-\frac{1}{2N}\sum_{k}\frac{\Gamma_{Q}^{(0)N}(q,k)\Gamma_{\omega Q}^{A}(k,p)}{E_{\frac{Q}{2}+k}+E_{\frac{Q}{2}-k}+\omega-i0^{+}}-\frac{1}{2N}\sum_{k}\frac{\Gamma_{Q}^{(0)A}(q,k)\Gamma_{\omega Q}^{N}(k,p)}{E_{\frac{Q}{2}+k}+E_{\frac{Q}{2}-k}-\omega-i0^{+}},\label{eq:BSeq2}
\end{align}
\end{widetext}where $\Gamma_{Q}^{(0)N/A}$ is the bare scattering
amplitude of the normal and anomalous type. The energy of the magnon pair bound state can be extracted from the
poles of the scattering amplitude. 
For a wide range of  $J_{2}/|J_{1}|$ values,
the bound state dispersion has its minimum at $Q=\pi$. The energy of the two-magnon  bound state
increases with $\alpha$, implying that the  critical field for the ``two-magnon'' condensation
decreases relative to the saturation field of the isotropic ($\alpha=0$) Hamiltonian. 

For the isotropic Heisenberg model, the condensate wave function has a
 $U(1)$ phase freedom. The spin anisotropy reduces this freedom to
 $Z_{2}$. The
Ginzburg-Landau theory tells us that the phase of the macroscopic wave
function can  either be real or imaginary depending on the sign of the
effective anomalous coupling parameter $u$. In
this section we  determine the phase of the wave function and also include a 
microscopic calculation of the parameter $u$. These properties are enclosed in the scattering amplitude
obtained from the solution of the Bethe-Salpeter equation. Our analysis 
shows that  wave function of the magnon pair condensate is
real, implying that the dominant order parameter is ${\cal O}^{a}$. 

\subsubsection{Wave function of the bound state}

We start  by introducing the two-magnon
Green's function, which can be easily obtained through
the scattering amplitude $\Gamma_{\omega Q}^{N/A}(q,p)$:
\begin{align}
G^{(2)}(\omega,Q;q,p) & =G_{0}^{(2)}(\omega,Q;,q,p)+\frac{1}{4}G_{0}^{(2)}(\omega,Q;q,q^{\prime})\nonumber \\
 & \times\Gamma_{\omega Q}^{N}(q^{\prime},p^{\prime})G_{0}^{(2)}(\omega,Q;p^{\prime},p),\label{eq:2pGFa}\\
G_{A}^{(2)}(\omega,Q;q,p) & =\frac{1}{4}G_{0}^{(2)}(-\omega,-Q;-q,-q^{\prime})\nonumber \\
 & \times\Gamma_{\omega Q}^{A}(q^{\prime},p^{\prime})G_{0}^{(2)}(-\omega,-Q;-p^{\prime},-p),\label{eq:2pGFb}
\end{align}
where $G_{0}^{(2)}$ is the non-interacting two particle Green's function:
\begin{equation}
G_{0}^{(2)}(\omega,Q;q,q^{\prime})=\frac{\delta(q-q^{\prime})-\delta(q+q^{\prime})}{\omega-E_{\frac{Q}{2}+q}-E_{\frac{Q}{2}-q}+i0^{+}}.
\end{equation}
The Lehmann representation shows explicitly that the two particle Green's function
has the following singular behavior near the pole of bound state, 
$\omega\sim\omega_{B}$:
\begin{align}
G_{N}^{(2)}(\omega,Q;q,p) & \sim\frac{\psi_{Q}(p)\psi_{Q}^{*}(q)}{\omega-\omega_{B}+i0^{+}}+\text{regular terms},\label{eq:singular1}\\
G_{A}^{(2)}(\omega,Q;q,p) & \sim\frac{\psi_{Q}(p)\phi_{-Q}(q)}{\omega-\omega_{B}+i0^{+}}+\text{regular terms},\label{eq:singular2}
\end{align}
where the regular terms come from higher excited states and $\omega_{B}>0$
is the bound state energy relative to the ground state. The projection of the bound state wave function on the two-magnon sector is obtained from the residue
of the pole,
\begin{align}
\psi_{Q}(p) & =\langle G|\alpha_{\frac{Q}{2}+p}\alpha_{\frac{Q}{2}-p}|B\rangle_{Q},\label{eq:bdwf1}\\
\phi_{\bar{Q}}^{*}(p) & =\langle G|\alpha_{-\frac{Q}{2}-p}^{\dagger}\alpha_{-\frac{Q}{2}+p}^{\dagger}|B\rangle_{Q},\label{eq:bdwf2}
\end{align}
where  $|B\rangle$ is the ket of the bound state and $|G\rangle$ the ground
state. Due to the anomalous interaction arising from the spin anisotropy term,
the bound state wave function is a linear combination of states with different particle number. 
The poles of the two particle Green's function given by Eqs.~(\ref{eq:2pGFa}) and (\ref{eq:2pGFb}) are obtained after
inserting the scattering amplitude, $\Gamma_{\omega Q}^{N/A}(q,p)$, which results from the  Bethe-Salpeter equation. According to Eqs.~($\ref{eq:bdwf1}$)
and (\ref{eq:bdwf2}), the bound state wave function is then obtained by extracting
the residue near the pole $\omega_{B}$. For $Q=\pi$,
the bound state wave functions, $\psi_{\pi}(p)$ and $\phi_{\pi}(p)$,
are even under spatial inversion. This result is found to be
correct for all  $J_{2}/|J_{1}|$ ratios and for any value of the bound state
energy $\omega_{B}$, indicating that the broken symmetry state below the critical field must 
preserve the spatial inversion symmetry.

\begin{figure}[H]
\centering
\includegraphics[scale=0.9]{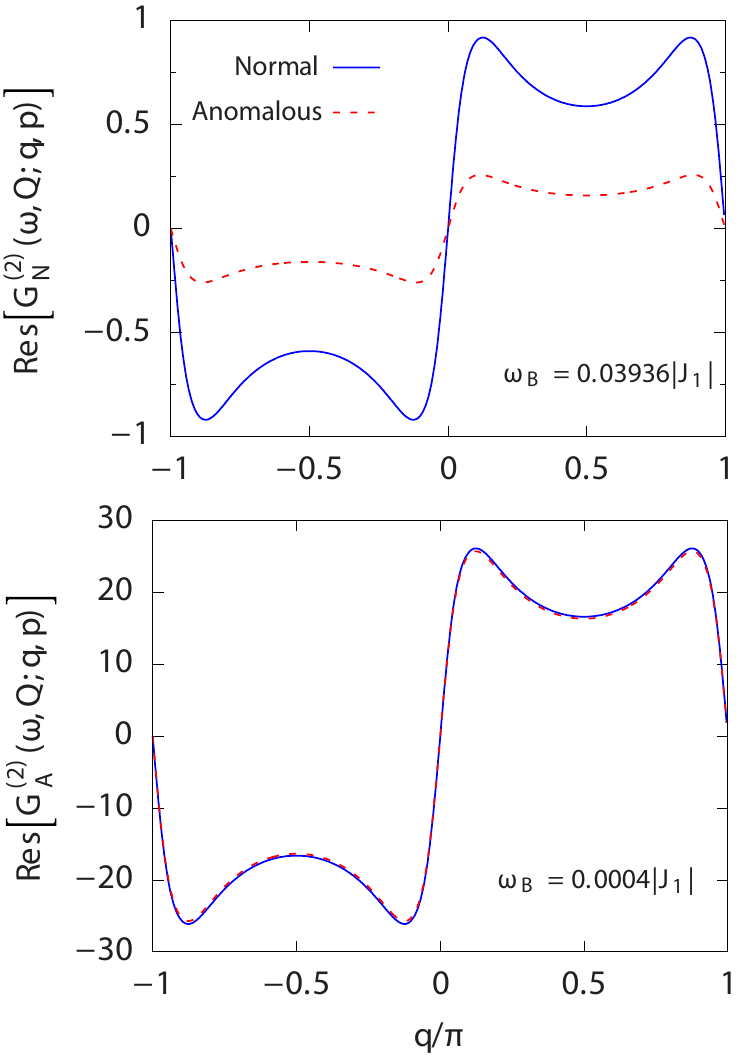}
\caption{Residue of the two particle normal/anomalous Green's function near the pole $\omega_{B}$
in frequency space for: (a)  $\omega_{B}/|J_{1}|=0.03936$ and (b)
 $\omega_{B}/|J_{1}|=0.0004$ (close to the ``two-magnon'' condensation point). 
 The other parameters are taken as $J_{2}/|J_{1}|=1,\alpha=0.2$,  momentum $p=0.9\pi$.
We find the normal (blue solid line) and anomalous (red dotted line)
two-particle Green's function become the same upon approaching
the critical condensation point.}
\label{fig:residue}
\end{figure}

As the system approaches the critical field corresponding to the onset of the ``two-magnon'' condensate, $\omega_{B}\rightarrow0$, the normal and the anomalous Green's function  become
exactly the same (see Fig. \ref{fig:residue}). Consequently,
the particle pair wave function, $\psi_{Q}(p)$, is exactly the same
as the hole pair wave function $\phi_{\bar{Q}}^{*}(p)$. This is
a manifestation of an emergent particle-hole symmetry at zero energy,
which sets a constraint on the phase of the bound state wave
function. If we describe the condensate state as a
coherent state built with the two-body bound state wave function,
the phases $\theta_{\pm}$ for the macroscopic components $\Psi_{\pm Q}$
are given by $\psi_{Q}(p)$ and $\phi_{\bar{Q}}(p)$, respectively.
The relationship $\psi_{Q}(p)=\phi_{\bar{Q}}^{*}(p)$ indicates that
$\theta_{+}+\theta_{-}=0$, which leads to a real order parameter
$\langle S_{r}^{+}S_{r+1}^{+}\rangle$ in real space.

Beyond the condensation point, the new ground state, $|G\rangle$, is characterized by the order parameter
 $\Phi_{Q}(q)=\langle G|\alpha_{\frac{Q}{2}+q}\alpha_{\frac{Q}{2}-q}|G\rangle$. 
 The coherent representation enables us to identify the bound state wave functions  $\psi_{Q}(p)$ and $\phi_{\bar{Q}}(p)$
with the two component order parameter, $\Phi_{Q}(p)$ and $\Phi_{\bar{Q}}(p)$, that was discussed 
in Section~\ref{GG}. This correspondence leads to the self-consistent
equation for the order parameter based on the Bethe-Salpeter equation,
from which one can straightforwardly confirm which order is favored. The analysis becomes more transparent by adopting
an equivalent but more straightforward approach. We just
introduce a small pairing field term into the Hamiltonian, which couples to the
order parameter: 
\begin{align}
{\cal H}_{paring} & =h_{Q}^{*}\sum_{q}\alpha_{\frac{Q}{2}+q}\alpha_{\frac{Q}{2}-q}+h_{\bar{Q}}^{*}\sum_{q}\alpha_{\frac{\bar{Q}}{2}+q}\alpha_{\frac{\bar{Q}}{2}-q}+h.c..
\end{align}
The renormalized pairing fields $h_{Q}^{R},h_{\bar{Q}}^{R}$ are indicated 
by the ladder series of vertex corrections in Fig.~\ref{fig:pairing}.
The pairing susceptibility diverges at the condensation point, implying that the
order parameter develops spontaneously beyond this point, i.e., 
in absence of the pairing fields $h_{Q}$ and $h_{\bar{Q}}$. The order
parameter in momentum space, $\Phi_{Q}(q)=\langle\alpha_{\frac{Q}{2}+q}\alpha_{\frac{Q}{2}-q}\rangle$,
 can be calculated as $\Phi_{Q}(q)=-h_{Q}^{R}(q)/\Omega_{q}$.
Therefore, the ladder series of vertex corrections in Fig. \ref{fig:pairing}
leads to the following self-consistent equation:
\begin{figure}[t!]
\centering
\includegraphics[scale=0.4]{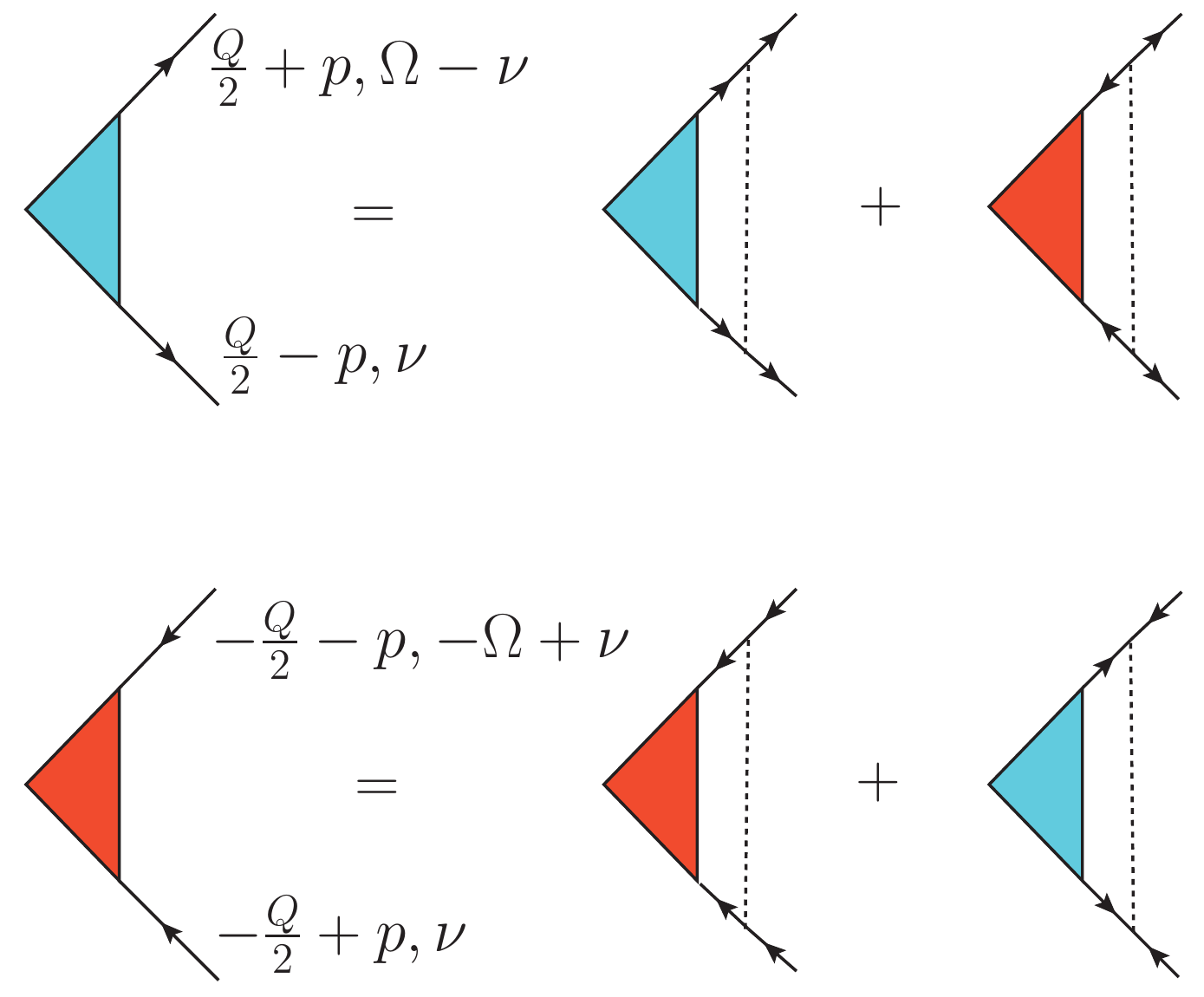}
\caption{Ladder diagrams for the pairing field $h_{\pm Q}^R(q)$. The blue (red) vertex represents the renormalized pairing field at
$Q$ ($-Q$). The dashed lines represent the symmetrized interaction vertexes in Eqs.~(\ref{eq:normalV}) and (\ref{eq:anomalousV}).}
\label{fig:pairing}
\end{figure}
\begin{align}
\sum_{k}\left(\Omega_{Q}(q)\delta_{q,k}+\frac{1}{2N}\Gamma_{Q}^{(0)N}(q,k)\pm\Gamma_{Q}^{(0)A}(k,q)\right)\Phi_{Q}^{\pm}(k)=0,\label{eq:eom}
\end{align}
where $\Omega_{Q}(q)=E_{{Q\over 2}+q}+E_{{Q\over 2}-q}$ is the energy of a two-magnon excitation and
\begin{equation}
\Phi_{Q}^{\pm}(k)=\Phi_{Q}(k)\pm\Phi_{-Q}^{*}(k).
\end{equation}
The order parameters $\Phi_{Q}^{\pm}(k)$ become finite when the corresponding matrix in Eq.~(\ref{eq:eom})
is singular. For $Q=\pi$,  $\Phi_{\pi}^{+}(k)$ coincides with
imaginary part of $\Phi_{\pi}(k)$, while $\Phi_{\pi}^{-}(k)$ is the
real part. The numerical calculation shows that
the order parameter $\Phi_{\pi}(k)$ is purely imaginary for $Q=\pi$  and 
that it satisfies $\Phi_{Q}(k)=\Phi_{-Q}^{*}(k)$ for $Q<\pi$.

To understand the meaning of this result in real space, we just need to
consider the order parameter $\langle S_{r+1}^{-}S_{r}^{-}\rangle$,
which is given by $\langle c_{r+1}c_{r}\rangle$ in terms of the Jordan-Wigner
fermionic annihilation operators. By applying Fourier and Bogoliubov transformations,
we find
\begin{align}
 & \langle S_{r+1}^{-}S_{r}^{-}\rangle=\Psi_{0}-\frac{1}{N}\sum_{q}u_{\frac{Q}{2}+q}u_{\frac{Q}{2}-q}\sin q\nonumber \\
 & \times[\Phi_{Q}(q)e^{i\frac{Q}{2}}e^{iQr}+\Phi_{\bar{Q}}(q)e^{i\frac{\bar{Q}}{2}}e^{i\bar{Q}r}]-\frac{1}{N}\sum_{q}v_{\frac{Q}{2}+q}v_{\frac{Q}{2}-q}\nonumber \\
 & \times\sin q[\Phi_{\bar{Q}}^{*}(q)e^{i\frac{Q}{2}}e^{iQr}+\Phi_{Q}^{*}(q)e^{i\frac{\bar{Q}}{2}}e^{i\bar{Q}r}].
\end{align}
The relationship $\Phi_{Q}(q)=\Phi_{\bar{Q}}^{*}(q)$ implies that the real space order parameter is
\begin{align}
\langle S_{r+1}^{-}S_{r}^{-}\rangle & =\Psi_{0}-\frac{2}{N}\sum_{q}\left(u_{\frac{Q}{2}+q}u_{\frac{Q}{2}-q}+v_{\frac{Q}{2}+q}v_{\frac{Q}{2}-q}\right)\nonumber \\
 & |\Phi_{Q}(q)|\sin q\cos\left(Qr+Q/2+\phi_{+}\right),
\end{align}
where $\phi_{+}$ is the phase  of $\Phi_{Q}(q)$. Therefore,
the order parameter $\langle S_{r+1}^{-}S_{r}^{-}\rangle$ is real.

\subsubsection{Microscopic calculation of phenomenological parameters}

\begin{figure*}
\includegraphics[width=\textwidth]{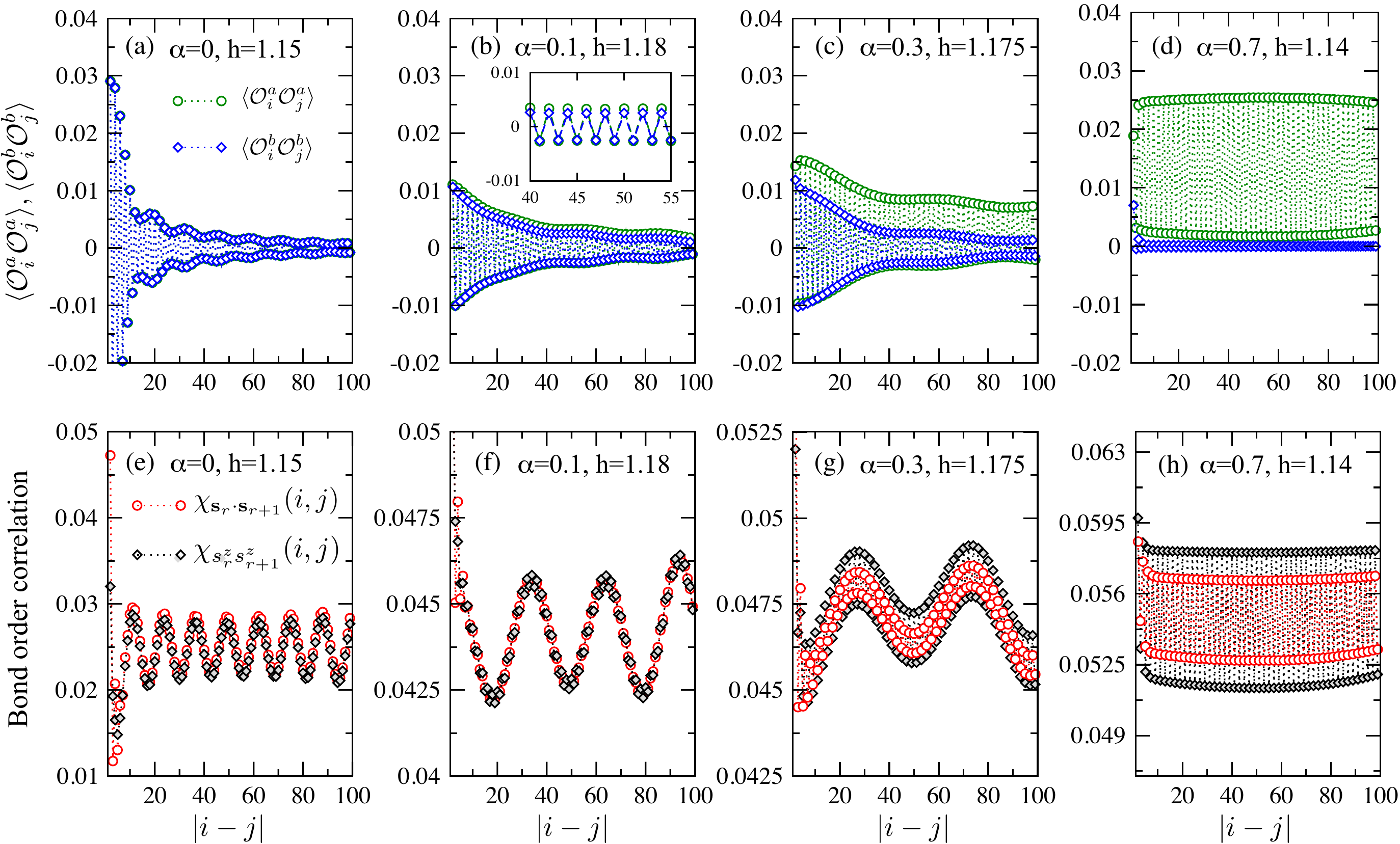}
\caption{(Color online) DMRG simulation for a spin chain of length $L=160$ with open boundary condition. The measurement is made 
in the bulk with $i=30$ and $31 < j \leq130$. (a)-(d) show the correlation function of two types
of bond order for different values of $\alpha$. (e)-(h) show the correlation function of the bond orders 
$ \chi_{{\bf S}_r \cdot {\bf S}_{r+1}}(i,j) = \langle {\bf S}_{i}\cdot{\bf S}_{i+1} {\bf S}_{j}\cdot{\bf S}_{j+1} \rangle$ and $ \chi_{ S^z_{r} S^z_{r+1}} (i,j) = \langle {S}^z_{i}{ S}^z_{i+1} {S}^z_{j}{ S}^z_{j+1} \rangle$.
The frustration ratio is $J_{2}/|J_{1}|=0.9659$ and the magnetic field is close to the critical field for each value of $\alpha$.}
\label{fig:dmrg}
\end{figure*}

To provide a microscopic derivation of the phenomenological parameters $r$ and $u$
we express the Ginzburg-Landau free energy in its diagonal form:
\begin{equation}
{\cal F}=\frac{r+u}{2}|\Psi_{+}|^{2}+\frac{r-u}{2}|\Psi_{-}|^{2},\label{eq:GLdiagonal}
\end{equation}
where $\Psi_{\pm }=\Psi_{Q}\pm\Psi_{\bar{Q}}^{*}$ and $\Psi_{Q}$
is the Fourier transform of the bond order parameter $\langle S_{r}^{-}S_{r+1}^{-}\rangle$.
The order parameter can be expressed in terms of the fermionic Bogoliubov quasi-particles, 
\begin{equation}
\Psi_{\pm }=-\frac{e^{iQ/2}}{\sqrt{N}}\sum_{q}\sin(q)(u_{\frac{Q}{2}+q}u_{\frac{Q}{2}-q}\pm v_{\frac{Q}{2}+q}v_{\frac{Q}{2}-q})\Phi_{\pm }(q),\label{eq:orderBogoliubov}
\end{equation}
where $\Phi_{\pm }(q)=\Phi_{Q}(q)\pm\Phi_{\bar{Q}}^{*}(q)$ and
$\Phi_{Q}(q)=\langle\alpha_{\frac{Q}{2}+q}\alpha_{\frac{Q}{2}-q}\rangle$. From
Eq. (\ref{eq:GLdiagonal}), we can identify the phenomenological parameters
with the inverse of the corresponding static susceptibilities:
\begin{equation}
r\pm u=\langle \Psi^{\dagger}_{\pm} \Psi_{\pm }\rangle^{-1}=\left(\chi_{\pm }\right)^{-1}.
\end{equation}
In other words, $\chi_{\pm}$ are the  response
functions to pairing fields that couple linearly to the $\Psi_{\pm }$ order parameters given in  Eq.~(\ref{eq:orderBogoliubov}):
\begin{align}
\chi_{\pm } & =-\frac{2}{N}\sum_{qp}\frac{\sin q\sin pB_{q}^{\pm}B_{p}^{\pm}}{\Omega_{Q}(q)\Omega_{Q}(p)}\left(\Gamma_{0Q}^{N}(q,p)\pm\Gamma_{0Q}^{A}(q,p)\right)\nonumber \\
 & +\frac{2}{N}\sum_{q}\frac{\sin^{2}q(B_{q}^{\pm})^{2}}{\Omega_{Q}(q)},
\end{align}
where $\Gamma_{0Q}^{N/A}(q,p)$ is the scattering amplitude at zero
frequency, $\Omega_{Q}(q)=E_{\frac{Q}{2}+q}+E_{\frac{Q}{2}-q}$ and
$B_{q}^{\pm}=u_{\frac{Q}{2}+q}u_{\frac{Q}{2}-q}\pm v_{\frac{Q}{2}+q}v_{\frac{Q}{2}-q}$.
The second term is the non-interacting susceptibility of the Bogoliubov
fermions, which is negligible near the critical point where both $\Gamma_{0Q}^{N}(q,p)$
and $\Gamma_{0Q}^{A}(q,p)$ diverge. The finite $u$  value arises from  the non-zero anomalous scattering
amplitude, $\Gamma_{0Q}^{A}(q,p)$, which forces the two susceptibilities  $\chi_{Q}^{\pm}$ to be different:
\begin{equation}
u=\frac{\left(\chi_{Q}^{+}\right)^{-1}-\left(\chi_{Q}^{-}\right)^{-1}}{2}.
\end{equation}
Numerically, we always find $u<0$ for different ratios of $J_{2}/J_{1}$, in agreement with our previous discussions.

\section{Numerical simulations}

\begin{figure*}[!t]
\includegraphics[width=0.7\textwidth]{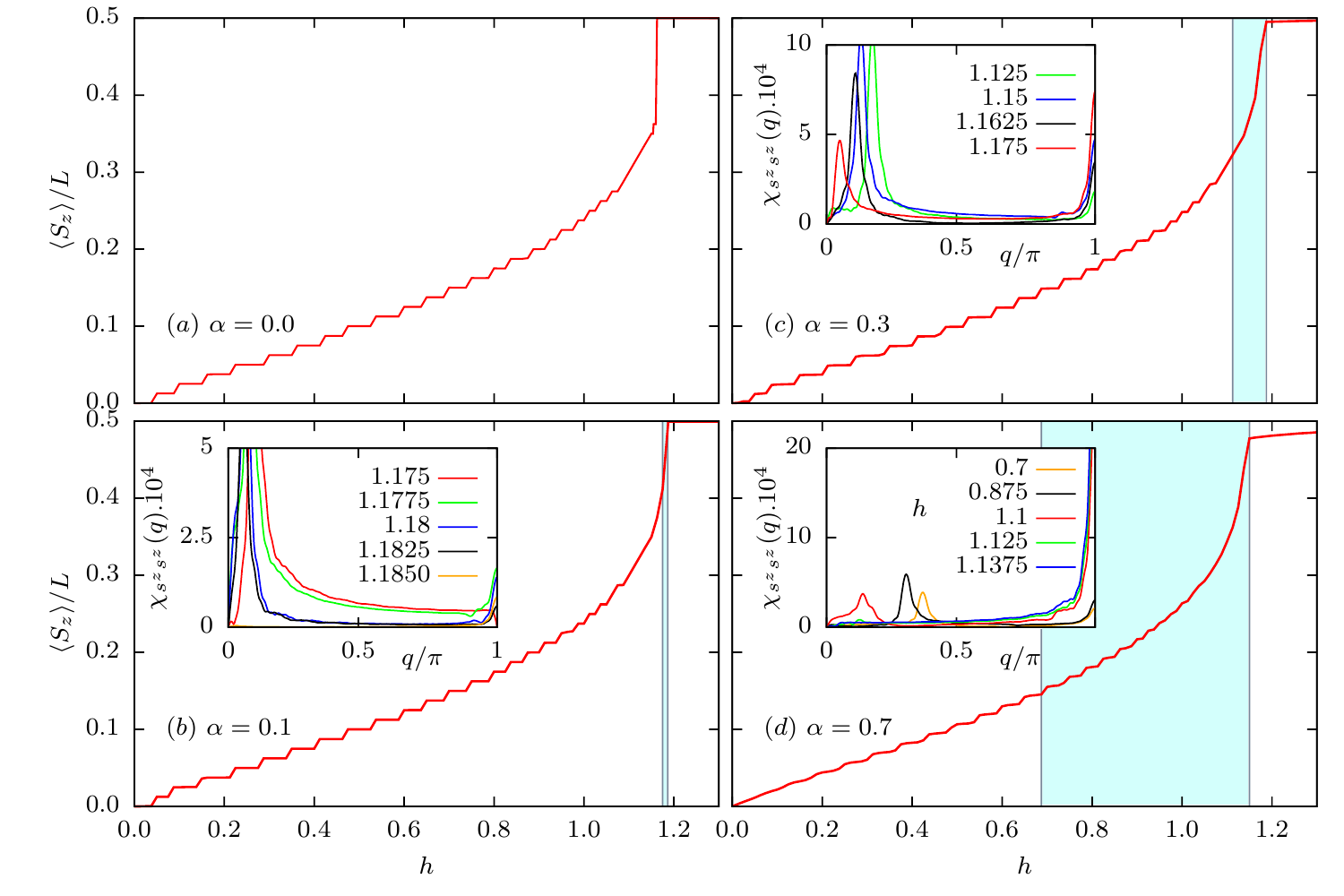}
\caption{(Color online) Magnetization curve obtained from DMRG simulations for the same parameters as in Fig.~\ref{fig:dmrg}.
The shaded (light blue) region indicates the extension of the dimerized phase.
The inset shows the Fourier transform of the real space correlation function $ \chi_{ S^z_{r} S^z_{r+1}} (i,j)$ for different magnetic field value. The peak at  $q=\pi$ is a clear indicator of bond dimerization arising from the combination of  
two-magnon condensation and spin anisotropy ($\alpha \neq 0$).
}
\label{fig:pipeak}
\end{figure*}

The above theoretical analysis for small $\alpha$ indicates that the spin anisotropy stabilizes a dimerized ground state.
In this section we present  DMRG calculations~\cite{ref:dmrg1,ref:dmrg2} for the anisotropic 1D  spin Hamiltonian ${\cal H}$, which confirm this analysis. The calculations have been done right below the critical field for chains of $L=160$ spins with open boundary conditions. 
We used up to 400 states and kept the truncation tolerance below $10^{-8}$ throughout the DMRG iterations. We  did 6 full sweeps of finite algorithm of DMRG to get well converged  observables.
Fig.~\ref{fig:dmrg} shows the correlation functions for the real and imaginary parts of the nematic  order parameter, ${\cal O}^{a}$, ${\cal O}^{b}$, 
for the ``pair-density" operator $S_i^z S_{i+1}^z$, and for the bond operators ${\bf S}_i \cdot {\bf S}_{i+1}$, for different values of $\alpha$.  The frustration ratio is taken as $J_2/J_1 = 0.9659$, which gives rise to a lowest energy ``two-magnon'' bound state with center of mass momentum equal to $\pi$.
In agreement with the ``two-magnon'' calculation, the correlation functions of the order parameters ${\cal O}^{a}$ and ${\cal O}^{b}$ oscillate with wave vector $\pi$. The long wave length oscillations are just a consequence of the open boundary conditions and the large spin density wave susceptibility. Note, however, that the incommensurate nature of oscillations precludes the possibility of having long range spin density wave ordering (the incommensurate spin density wave ordering breaks the continuous U(1) symmetry group of translations).

The first column in Fig.~\ref{fig:dmrg} with $\alpha=0$ corresponds to  the U(1) symmetric case for which all the correlators exhibit the expected power law behavior. The order parameters ${\cal O}^{a}$ and ${\cal O}^{b}$ are connected by a $\pi/4$ spin rotation about the $z$-axis, which is a symmetry of the Hamiltonian for $\alpha=0$. Correspondingly, both
correlation functions exhibit an identical power-law decay. 
The bond and the pair density correlators also exhibit a power law decay with long wave length oscillations, which are magnified by the open boundary conditions~\cite{ref:chiral2}.

The upper panels of Fig.~\ref{fig:dmrg} show that the real component of the nematic order parameter dominates over the imaginary part and develops long range ordering for  nonzero $\alpha$. The amplitude of the  order parameter increases with $\alpha$. As expected from the previous analysis, the pair density, $S_i^z S_{i+1}^z$, and the bond, ${\bf S}_i \cdot {\bf S}_{i+1}$, operators  become dimerized as a consequence of the coexistence of uniform and staggered components of the ${\cal O}^a$ order parameter. We recall that the uniform component is directly induced by the $\alpha$ term of the Hamiltonian, while the staggered ($\pi$) component is spontaneously generated. 
The dimerization   is identified by Fourier transforming the real space correlation functions shown in Figs.~\ref{fig:dmrg}(e-h). This information is provided  in the insets of Fig.~\ref{fig:pipeak}, which show a clear peak at $Q=\pi$.
It is also clear from Figs.~\ref{fig:dmrg}(e-h) that the dimerization order parameter becomes more robust upon increasing $\alpha$. 
Moreover, the insets of Fig.~\ref{fig:pipeak}  show that the Friedel oscillations of $\langle S_i^z \rangle$ are strongly suppressed for  $\alpha = 0.7$, i.e., for a large amplitude of the dimerization order parameter. Consistently with these results,
Fig.~\ref{fig:pipeak}  shows that  the dimerized phase (shaded region) becomes stable over a larger  window of magnetic field values
as we increase $\alpha$.  The same figure includes the field dependence of the magnetization curve for different values of $\alpha$. As expected, the slope of the  magnetization curve above the critical field increases with $\alpha$  and the magnetization saturates  asymptotically for $H\rightarrow \infty$.

\section{Discussion and summary}

Finally, we discuss the consequences of our theoretical study on the  experimental search for  nematic phases in various quasi-1D materials. We first note that two previous works studied the role of an XXZ magnetic anisotropy, \emph{which preserves} the $U(1)$ symmetry of the spin Hamiltonian. 
Both works concluded that the effect of the U(1) invariant spin anisotropy is to stabilize the nematic phase. Similarly, we are finding that the effect of a
U(1) symmetry breaking anisotropy is to stabilize the dimerized phase, which is a ``direct descendant" of the nematic ordering (real part of the nematic order parameter).

The spin anisotropy of the  quasi-1D compound LiCuVO$_4$ is  approximately $\alpha = 0.1$.~\cite{ref:Z2material1, ref:1Dmaterial2, ref:1Dmaterial3, ref:1Dmaterial6,LiCuVO4,LiCuVO4b, observation}
The results of our simulations for this value of $\alpha$ [see Figs.~\ref{fig:dmrg} (b) and (f)] clearly show that the bond-bond correlation function develops a visible $\pi$ ordering (dimerization).  This is a salient experimental signature that can be detected with X-rays.
Rb$_2$Cu$_2$Mo$_3$O$_{12}$ is another quasi-1D   $J_1$-$J_2$ frustrated magnet \cite{ref:1DmaterialA1,ref:1DmaterialA2} with $J_2/\lvert J_1 \rvert \simeq 0.33$. This value falls in the ``mutipolar" phase (quasi-condensate of $n$-magnon bound states with $n>2$) of  the U(1) invariant spin model. Indeed, the nematic phase becomes unstable for  $J_2/\rvert J_1 \rvert <0.37$. Although we have not considered the situation of multipolar orderings of higher order than nematic, our simple GL analysis shows that the $\alpha$-term  should once again select the real or the imaginary part of the multipolar order parameter. 

A more recent experimental work reports evidence of  a gapped ``nematic ordering" for the quasi-1D $J_1-J_2$ compound LiCuSbO$_4$.~\cite{grafe2016} Nuclear magnetic resonance (NMR) measurements indicate the opening of a rather large spin gap above a critical magnetic field value $H_{c1} \simeq 13$T. Conventional explanations for the origin of this spin gap, such as saturation of the magnetization  or the presence of a staggered  Dzyaloshinskii-Moriya (DM) interaction have been ruled out by the experiments. Moreover, the absence of a kink in the low temperature  magnetization vs. field curve is consistent with an  XYZ magnetic anisotropy. Based on our analysis, this U(1) symmetry breaking anisotropy is responsible for the the spin gap that is inferred from the NMR measurements. According to Ref.~\cite{grafe2016} the value of the spin anisotropy is $\alpha \simeq 0.08$. In addition, a $J_2/\rvert J_1 \rvert \simeq 0.28 $ ratio is obtained  from  a fit of  the magnetization curve with a pure one-dimensional Hamiltonian. 
This ratio puts the material outside the region of stability of the nematic phase~\cite{ref:chiral2} (higher multipolar orderings are expected for $J_2/\rvert J_1 \lesssim 0.367$). However, Ref.~\cite{LiCusbO4} reports a significantly larger ratio $J_2/\rvert J_1 \rvert \simeq 0.45$, which falls in the nematic phase of the U(1) invariant Hamiltonian.~\cite{ref:chiral2}. If this   $J_2/\rvert J_1 \rvert$ ratio is correct,  the gapped high-field spin phase of this compound must correspond to a spin dimerized phase. 
However, further inspection of this material~\cite{grafe2016} shows that the lattice is already dimerized (consecutive bonds are not equivalent) implying a continuous crossover between the dimerized spin phase and the high field paramagnetic state.


In summary, we have demonstrated that the spin-orbit interaction has important consequences for the 
field-induced spin nematic ordering of U(1) invariant frustrated models.
The symmetry reduction of ${\cal H}$ due to the presence of the  Ising term renders the  nematic order parameter 
inapplicable. However, the real and imaginary parts of the nematic bond order parameter still break  discrete symmetries, which can be directly related with   observable quantities.
Our analytical and numerical results demonstrate that the  spin-orbit interaction stabilizes a bond density wave
(bond dimerization for $Q=\pi$) which couples to the lattice via the magneto-elastic effect.
These results are confirmed by our DMRG simulations. Given that the spin-orbit interaction is ubiquitous in nature and that continuous symmetries are never strictly present in real magnets, our study indicates that the proposed nematic ordering is likely to be replaced by bond dimerization in systems described by  a $J_1$-$J_2$ model with $|J_2|/|J_1| \gtrsim 0.38$. Even in quasi-1D systems, which are approximately described by a U(1) invariant $XXZ$ model, the application of a magnetic field  \emph{perpendicular} to the chains should induce the dimerized state that we are proposing here. 

\begin{acknowledgments}

We thank Zhentao Wang, Shi-Zeng Lin and Yukitoshi Motome for helpful discussions.
S. Z. and C.D.B. are supported by funding from the Lincoln Chair of Excellence
in Physics and from the Los Alamos National Laboratory
Directed Research and Development program. 
N. K. was supported by the National Science Foundation, under Grant No. DMR-1404375.
E. D. was supported by the U.S. Department of Energy, Office of Basic Energy Sciences, Materials
Sciences and Engineering Division.

\end{acknowledgments}

\end{document}